\documentclass[10pt,twoside,BCOR7mm,DIV15,headinclude,footexclude,cleardoubleempty,idxtotoc]{scrartcl}

\usepackage[english]{babel}
\usepackage{graphicx}
\usepackage{hyperref}
\usepackage{scrpage2}
\usepackage{hyperref}
\usepackage{ifthen}

\makeatletter
\renewcommand{\@biblabel}[1]{}
\renewcommand{\@cite}[2]{%
{#1\ifthenelse{\boolean{@tempswa}}{,#2}{}}}
\makeatother

\hypersetup{breaklinks=true
,colorlinks=true,linkcolor=black,urlcolor=blue
,citecolor=black}

\ofoot{\thepage}
\ifoot{}

\setheadsepline{1pt}

\setkomafont{pagehead}{\normalfont\sffamily}
\setkomafont{pagenumber}{\normalfont\rmfamily}

\usepackage{booktabs}
\usepackage{amsmath}
\usepackage{amssymb}
\usepackage{multicol}
\usepackage{float}

\makeatletter
\newcommand{\listofcontributions}{\@starttoc{con}}

\newcommand{\l@contribution} {\@dottedtocline{1}{1.5em}{2.3em}}
\makeatother

\newenvironment{contribution}{
\setcounter{section}{0}
\setcounter{figure}{0}
\setcounter{table}{0}
\begin{flushleft}
{\em Clumping in Hot Star Winds \\
W.-R.\ Hamann, A.\ Feldmeier \& L.\ Oskinova, eds.\\
Potsdam: Univ.-Verl., 2007 \\
URN: http://nbn-resolving.de/urn:nbn:de:kobv:517-opus-13981
} 
\end{flushleft}
}{
\newpage
\lehead{}
\rohead{}
}

\begin{document}

\setlength{\baselineskip}{2.5ex}

\begin{contribution}

\lehead{M. C.\ Runacres}

\rohead{Hydro models of outer-wind clumping}

\begin{center}
{\LARGE \bf Hydrodynamical models of clumping beyond $50~R_{*}$}\\
\medskip

{\it\bf M. C. Runacres}\\

{\it Erasmushogeschool Brussel, Belgium}\\

\begin{abstract}
We present one-dimensional, time-dependent models of the
clumps generated by the line-deshadowing instability. In order
to follow the clumps out to distances of more than $1000~R_*$, 
we use an efficient moving-box technique. We show that, within the
approximations, the wind can remain clumped well into the
formation region of the radio continuum.
\end{abstract}
\end{center}

\begin{multicols}{2}

\section{Introduction}
The line-driven stellar winds of hot stars are subject to a strong
line-deshadowing instability (e.g. Owocki \& Rybicki \cite{runacres:OR84}),
which causes the wind
to become highly structured.
This structure takes the form of strong shocks, strong density contrasts
and regions of hot, but generally rarefied, gas.

The structure caused by the line-deshadowing instability
is small-scale and stochastic in nature, as opposed to
the large-scale, coherent structure associated with discrete absorption 
components and related features in ultraviolet spectral lines of hot stars 
(Prinja \cite{runacres:Prinja98})
We use the word {\it clumping}
to refer to the small-scale density structure only, 
with the line-deshadowing instability as its most likely cause.

The degree of clumping at a certain distance $r$ from the star is most
readily described by the clumping factor $f_{\rm cl}$, defined
as
\[ f_{\rm cl} (r)= \frac{<\rho^2>}{<\rho>^2},
\]
where $<>$ denotes the time-averaging of the quantity between brackets.
If all of the mass is concentrated in the dense clumps, then
the clumping factor is the inverse of the volume filling factor, and 
is equal to the overdensity of the clumps with respect to the mean wind:
\[f_{\rm cl} = \frac{1}{f_{\rm volume}} = \frac{\rho_{\rm clump}}{<\rho>}.
\]
The mass-loss rate derived from a density-squared dependent
observational diagnostic
is inversely proportional to the square root of the clumping factor.

Most theoretical studies of clumping are limited to the wind below
30 stellar radii ($R_*$). There is, however, ample reason to
study clumping at much larger distances from the star. The radio continuum
used to derive the mass-loss rates of hot stars
is formed by free-free emission and
hence is strongly sensitive to clumping. 
To know the true value of the mass-loss rate, we therefore need to
know the degree of clumping.  The same holds for other diagnostics
of the mass-loss rate that are proportional to the density squared, 
such as $H\alpha$.
As is shown throughout these proceedings, 
this is a surprisingly difficult thing to do.
All mass-loss rate diagnostics are affected by uncertainties. One
way to reduce such uncertainties, is to combine different observational
mass-loss rate diagnostics, formed in different parts of the wind, 
to obtain the radial stratification of clumping. Such a study
has been performed by Puls et al. (\cite{runacres:Puls+2006}).

Even from such a study, it is impossible to derive absolute values 
of the clumping factor. 
If one derives a certain radial stratification of the clumping factor assuming
the clumping vanishes in the radio formation region, then 
the observations can also be explained by this clumping factor multiplied 
by a constant factor, providing the mass-loss rate is lowered accordingly.
The derived value of the clumping factor 
(and hence the value of the mass-loss rate) thus depends on the assumption
one makes about the amount of clumping in the radio formation region. 
Therefore it is important to gain insight in 
how clumps evolve as they move out to large distances, 
and to investigate whether clumps can survive as far as
the radio formation region.

\section{Hydrodynamical models}
\subsection{Hydrodynamical models including the line-deshadowing instability}
We solve the conservation equations of hydrodynamics, using the time-dependent
hydrodynamics code VH-1, developed by J. M. Blondin, and
modified by S.~P. Owocki to include the
acceleration due to line driving. 
Our models are one-dimensional. 
The radiative acceleration is included in the model
using the smooth source function method (Owocki \cite{runacres:SSF}). 
The structure is self-excited, in the sense
that there are no external perturbations at the base of the wind.
The structure is seeded by internal base perturbations, that arise
as radiation is scattered back to the base from the structured outer wind.
(Initial structure arises as the wind solution adapts from the smooth initial 
condition). In the absence of detailed knowledge of the photospheric 
perturbations acting at the base of a real wind, self-excited structure
can be seen as a conservative estimate of wind structure.

Radiative and adiabatic cooling
are included in the energy equation. Photo-ionisation heating is mimicked
by imposing a distance-dependent floor temperature, below which the 
temperature is not allowed to drop. 
Details are given in Runacres \& Owocki
(\cite{runacres:RO2002}). From test calculations performed in that paper, we 
have learned that the amount of
clumping depends on the value adopted for the floor temperature, as well
as on the strength of the line-driving. Also, it is necessary to maintain
a rather fine spacing of the radial mesh, in order to adequately 
resolve the structure. On the other hand,
clumping does not depend on the radiative force beyond 
$30~R_*$. This reduces the outer-wind evolution to a pure gasdynamical problem,
allowing us to construct vastly more economical models, which will be
presented in the next section. 

\subsection{Moving-box models}
For a star like $\zeta$~Pup, about 
half of the radio continuum is formed beyond $100~R_*$. So in order
to make meaningful predictions about the effect of clumping on the
radio mass loss rate, we need to model structure out to
very large distances from the star. 
Even without the evaluation of the radiative force, 
evolving the entire stellar wind (between 1 and say $1000~R_*$)
at the required high spatial resolution
is still very expensive. A solution is suggested by realising 
that the structure generated by the instability, apart from being
stochastic, is also quasi-regular in the sense that similar features are
repeated over time. 
Therefore it is not necessary to keep track of the 
whole stellar wind during the duration of the simulation. It is enough 
to select a limited but representative portion of the structure, 
and follow this ``box'' as 
it moves out at the terminal speed. Following a portion of the wind entails
transforming the conservation equations to a moving reference frame. This is
not possible directly, as the spherical equations of hydrodynamics are not
invariant under a Galilean transformation. This problem can be circumvented by
rewriting the equations in a pseudo-planar form. In this form, the equations
resemble the planar equations of hydrodynamics, while still describing
a spherical geometry. We impose periodic boundary conditions on the box, i.e.
structures that flow out of the box on one side, are made to enter it on
the other side. Details can be found in Runacres \& Owocki 
(\cite{runacres:RO2005}).

In the following section, we use a periodic box model, starting from
a hydrodynamical model including the line-deshadowing instability, to
predict the radial stratification of wind clumping. 
The adopted model parameters are the same as in  Runacres \& Owocki
(\cite{runacres:RO2005}).

\section{Results}
Fig.~\ref{runacres:fig2} shows the density contrast (density divided by mean
density) within the box
as a function of radius and time, as the box moves out from 
$\sim 100$ to $\sim 1300~R_*$. The backward running streaks are shells that are
somewhat slower than the terminal speed, the forward running streaks are
faster than the terminal speed. The streaks broaden as they evolve, reflecting
the fact that shells expand (at a few times the sound speed) as they move
out. Within the assumptions of the model, the clumpiness is maintained by
collisions between shells. As shells collide, they form denser shells, 
counteracting their pressure expansion.

\begin{figure}[H]
\begin{center}
\includegraphics[bb=30 140 580 680, clip, width=\columnwidth]{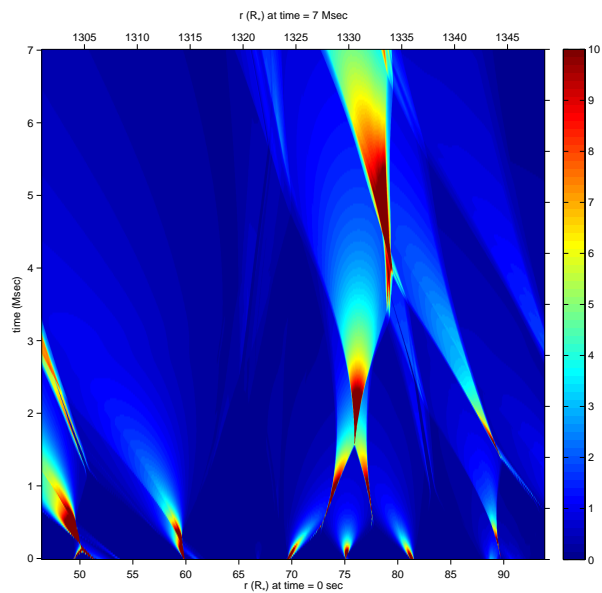}
\caption{Image of the density contrast  as a function 
of space and time. The intensity scale has been truncated to highlight the
kinematics of the shells.
The adopted model parameters are the same as in  Runacres \& Owocki 
(\cite{runacres:RO2005}).
\label{runacres:fig2}}
\end{center}
\end{figure}

The clumping factor for a moving-box model extending out 
to $\sim 1300~R_*$ is shown
in Fig.~\ref{runacres:fig1}. 
\begin{figure*}[!t]
\begin{center}
\includegraphics
  [width=0.99\textwidth]{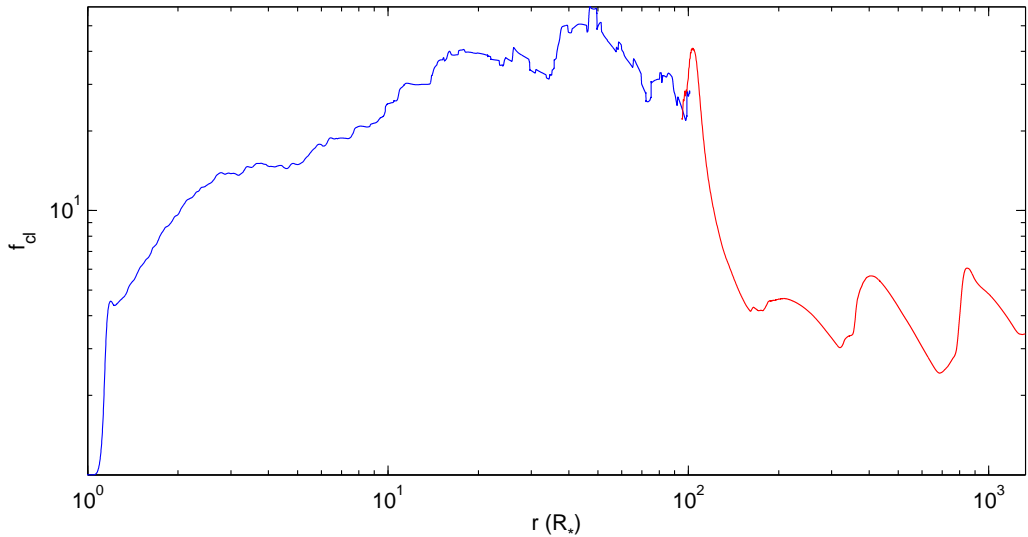}
\caption{The clumping factor as a function of radius, predicted by our
theoretical models. Below 100 $R_*$ the model is a line-driven instability 
model, above $100~R_*$ it is a moving-box model.
\label{runacres:fig1}}
\end{center}
\end{figure*}
Below 100 $R_*$ the model is a 
line-driven instability 
model, above $100~R_*$ it is a moving-box model. It is clear that these models predict that the winds stays clumped well into
the radio formation region, with clumping factors beyond $200~R_*$ ranging
from 2.5 to 6. Inferred mass-loss rates would therefore be overestimated by
a factor of two. 

\section{Discussion and conclusions}
Our models predict an increase of the clumping factor from the base of the wind
to $\sim 50~R_*$ (Fig.~\ref{runacres:fig1}),
after which the clumping factor decreases, maintaining a level of
residual clumping beyond $200~R_*$. 
This does not quite match the clumping
factors derived from observations by Puls et al. (\cite{runacres:Puls+2006}), 
which start to tail off closer to the star ($\sim 10~R_*$).
As has been mentioned before, the observations do not
tell us whether or not there is residual clumping at very
large distances from the star.

There are of course a number of limitations to our model. As has been mentioned
above, we have used self-excited structure without external perturbations.
Also, we have not attempted to model different spectral types. In particular,
the important difference between dense and less dense winds found by
Puls et al. (\cite{runacres:Puls+2006}) has not been investigated at all
in our models.

A key limitation of the present model is of course
its restriction to just one 
dimension. The focus here is entirely on the extensive radial structure, 
and the instabilities that are likely to break up the azimuthal coherence
of the structure are not accounted for. It remains to be seen to what extent
this changes the global evolution of instability-generated wind structure.
We plan to extend the present models to 2-D in the near future.


\end{multicols}

\end{contribution}


\end{document}